\documentclass[prl,twocolumn, reprint, superscriptaddress]{revtex4-1}
\usepackage{graphicx}
\usepackage{amsmath}
\usepackage{subfig}
\usepackage[justification=justified]{caption}

\newcommand{\G}{\mathbf{G}}
\newcommand{\rr}{\mathbf{r}}

\newcommand{\q}{\mathbf{q}}

\newcommand{\mr}{\mathrm}
\newcommand{\mb}{\mathbf}
\newcommand{\ve}{\mathbf{v}_e}
\begin{document} 

\title{Plasmons on the Edge of MoS$\mathbf{{}_2}$ Nanostructures}

\author{Kirsten Andersen}
\email{kiran@fysik.dtu.dk}
\affiliation{Center for Atomic-scale Materials Design, Department of
Physics \\ Technical University of Denmark, DK - 2800 Kgs. Lyngby, Denmark}

\author{Karsten W. Jacobsen}
\affiliation{Center for Atomic-scale Materials Design, Department of
Physics \\ Technical University of Denmark, DK - 2800 Kgs. Lyngby, Denmark}

\author{Kristian S. Thygesen}
\affiliation{Center for Atomic-scale Materials Design, Department of
Physics \\ Technical University of Denmark, DK - 2800 Kgs. Lyngby, Denmark}
\affiliation{Center for Nanostructured Graphene  \\ Technical University of Denmark, DK - 2800 Kgs. Lyngby, Denmark}

 
\begin{abstract}
  Using ab-initio calculations we predict the existence of
  one-dimensional (1D), atomically confined plasmons at the edges of a
  zigzag $\mathrm{MoS_2}$ nanoribbon. The strongest plasmon originates
  from a metallic edge state localized on the sulfur dimers decorating
  the Mo-edge of the ribbon. A detailed analysis of the dielectric
  function reveals that the observed deviations from the ideal 1D
  plasmon behavior result from single-particle transitions between the
  metallic edge state and the valence- and conduction bands of the
  MoS$_2$ sheet. The Mo- and S-edges of the ribbon are clearly
  distinguishable in calculated spatially resolved electron energy
  loss spectrum owing to the different plasmonic properties of the two
  edges. The edge plasmons could potentially be
  utilized for tuning the photocatalytic activity of MoS$_2$
  nanoparticles.
\end{abstract}

\maketitle
Plasmons are collective electronic excitations that couple strongly to
external fields such as light or fast propagating charges. In metals, their
excitation gives rise to charge oscillations which can be localized
either in the interior of the metal a or at its surface. Localized surface plasmon resonances (LSPR) in metal nanoparticles have been utilized
for chemical sensing \cite{kneipp_1997}, cancer treatment
\cite{au_2008} and to increase solar cell performance
\cite{atwater2010plasmonics}. Two-dimensional (2D) plasmons are found
at metal surfaces and thin films \cite{Pitarke_2007}, in
graphene \cite{grapheneplas2, grigorenko2012graphene, wunsch2006dynamical, hwang2007dielectric}, and other atomically thin 2D materials
such as metallic transition metal dichalcogenides \cite{TMDs}.

Compared to their 2D counterparts one-dimensional plasmons have been
much less studied. So far, true atomically
confined 1D plasmons have only been observed in self-assembled
atomic chains on semi-conducting substrates \cite{nagao2006one,
  krieg2013one}. One peculiar fundamental aspect of 1D metals is
the break down of Fermi-liquid theory and the transition to Luttinger
liquid behavior \cite{haldane1981luttinger,
  voit1995one}: At low energies the spectral weight of the
quasiparticle peak approaches zero and the life-time broadening
becomes of the same order as the excitation
energy \cite{yacoby1996nonuniversal, bockrath1999luttinger}. This
implies that for (ideal) 1D metals, the plasmons adopt a special
status as the only excitations which can couple to external fields.

Recently, the interest in novel 2D materials, such as graphene and
monolayer $\mathrm{MoS}_2$, has been accelerated by the prospects of
utilizing their unique electronic properties for nanoscale
(opto)electronics \cite{mak2010atomically, radisavljevic2011single}.
While graphene has shown great potential as a material for terahertz
plasmonics \cite{grapheneplas1}, the finite band gap of MoS$_2$ makes
it unsuitable for applications within this field. In a rather
different context, MoS$_2$ nanoparticles are used industrially as
hydrodesulphurization catalysts \cite{chianelli2006catalytic} and are
considered as promising alternative to Pt as low-cost catalysts for
hydrogen evolution \cite{jaramillo2007identification}. The high
catalytic activity of the MoS$_2$ nanoparticles has been directly
linked to the presence of metallic edges states on the otherwise
semi-conducting MoS$_2$ nanoparticles \cite{lauritsen2003chemistry}.
Very recently, an observed extraordinary photo-catalytic behavior of
metallic nanoparticles has been shown to originate from the excitation
of LSPRs \cite{christopher2012singular,marimuthu2013tuning}. As we
predict here, the the metallic edge states on MoS$_2$ nanostructures
can lead to the formation of highly localized 1D plasmons -- a result
which adds edge-plasmonics to the list of this material's unique
properties with potential applications within nanoplasmonics and
photocatalysis.

\begin{figure}
\begin{center}
  \includegraphics[width=\linewidth]{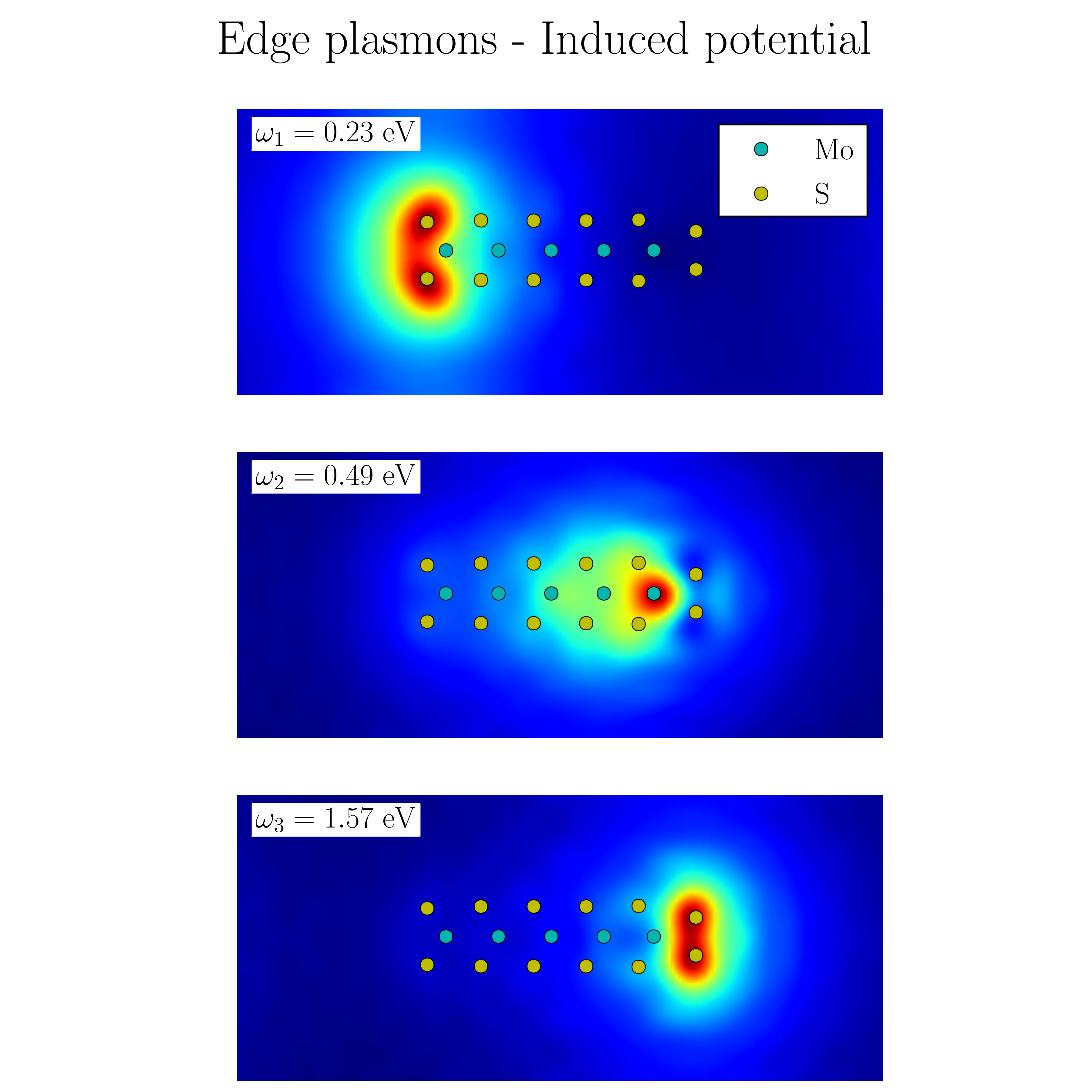}
\caption{Induced electric potential of the three plasmon eigenmodes at $q = 0.2 \mathrm{\AA}^{-1}$, calculated as the eigenvectors of $\epsilon_q(\mb{r},\mb{r}',\omega)$. The structure of the ribbon is shown in front view, with the ribbon running into the plane of the page. The Mo- and S-edge, covered by sulfur dimers, is found to the right and left of the ribbon respectively. }
\label{modes}
\end{center}
\end{figure}

In this communication, we use time-dependent density functional theory (DFT)
to demonstrate the existence of a set of highly localized plasmons on
the edges of an MoS$_2$ nanoribbon, see Fig.~\ref{modes}. The
fundamental properties of this new type of edge plasmon are
investigated through a spectral analysis of the dielectric function
which allows us to identify the plasmonic eigenmodes of the system in
an unambiguous way. The deviations of the edge plasmon properties from
those of an ideal 1D plasmon are shown to arise from interband
transitions between the metallic edge state and the MoS$_2$
conduction- and valence bands. The distinctly different plasmonic
properties of the Mo- and S-edges are clearly seen in the spatially
resolved loss spectrum suggesting that high resolution transmission
electron microscopy could be used to probe the atomic structure of
nanoribbon edges.

All calculations were performed with the GPAW electronic structure
code, which is a grid-based implementation of the projector-augmented wave method \cite{GPAW2}. 
We considered a 1.5 nm wide sulfur terminated zigzag
MoS$_2$ nanoribbon, see Fig.~\ref{modes} for a front view of the structure, and supplemental materials for a detailed image of the structure. 
This edge structure should be favorable at high chemical potential of sulfur, which can be tuned
by the choice of sulfiding agent \cite{bollinger2003atomic}. The
ribbon was placed in a supercell including 20 and 12 \AA~of vacuum in
the two perpendicular directions. The wave functions were represented
in a plane-wave basis set with an energy cutoff of 340 eV. The
structure was relaxed using the Perdew-Burke-Ernzerhof (PBE)  exchange-correlation functional,
while the GLLB-SC functional, including GLLB type exchange by Gritsenko
et al. and PBEsol correlation \cite{GLLBSC}, was applied to calculate
the single-particle states used as input for the linear response
calculation. The GLLB-SC functional has been shown to improve the PBE
band gaps of semi-conductors \cite{GLLBSC} and yield better
descriptions of plasmon resonances in noble metals \cite{yan2011first}
due to its improved description of the $d$-band energies. However, for
the MoS$_2$ nanoribbon GLLB-SC gives only minor corrections to the PBE
band energies.  The 1D Brillouin zone was sampled with 128 k-points
along the ribbon axis. A spin polarized calculation yielded zero
magnetic moment for the structure.

The dielectric matrix was calculated in the random phase approximation (RPA) following reference~\cite{Yan_2011}: 
\begin{equation}\label{eq:eps}
 \epsilon_{\mb{G},\mb{G}'}(\mb{q},\omega) = \delta_{\mb{G},\mb{G}'} - \frac{4\pi}{|\mb{q}+\mb{G}|^2} \chi^0_{\mb{G},\mb{G}'}(\mb{q},\omega).
\end{equation}
For the calculation of the non-interacting density response-function, $\chi^0$, empty states were included up to 12 eV above the Fermi level, which was sufficient to converge the low energy plasmon peaks of interest. An energy-cutoff of 30 eV was used for the reciprocal lattice vectors $\G$ and $\G'$. We used a non-linear frequency grid from 0 to 12 eV, with an initial grid-spacing of 0.01 eV at $\omega = 0$ and a smearing of 0.02 eV. The dense frequency sampling and small broadening was necessary at low energies in order to resolve all intra-band transitions. The Wigner-Seitz truncated Coulomb approximation \cite{sundararaman2013regularization} was used in order to avoid interaction between supercells, which would otherwise give a significant energy shift at low momentum transfers. 

\begin{figure}
\centering
\subfloat[\label{bands}{}]{\includegraphics[height=0.7\linewidth]{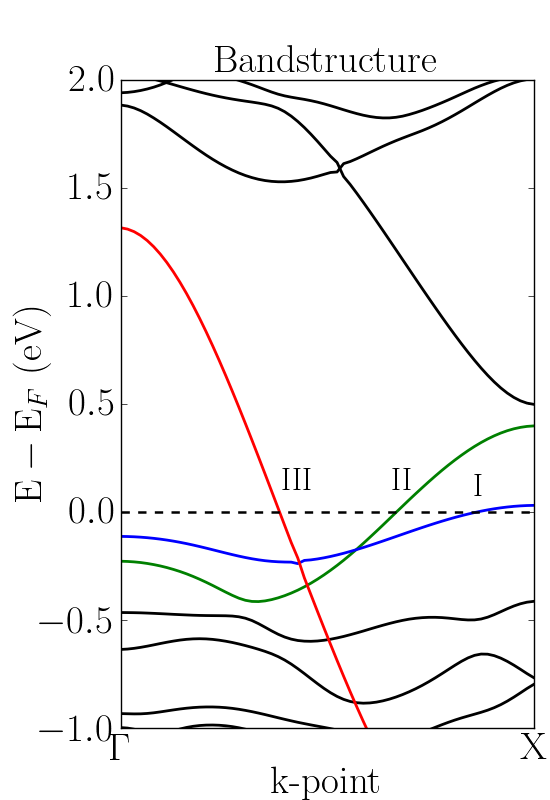}}
\hspace{5mm}
\subfloat[\label{eigs}{}]{\includegraphics[height=0.7\linewidth]{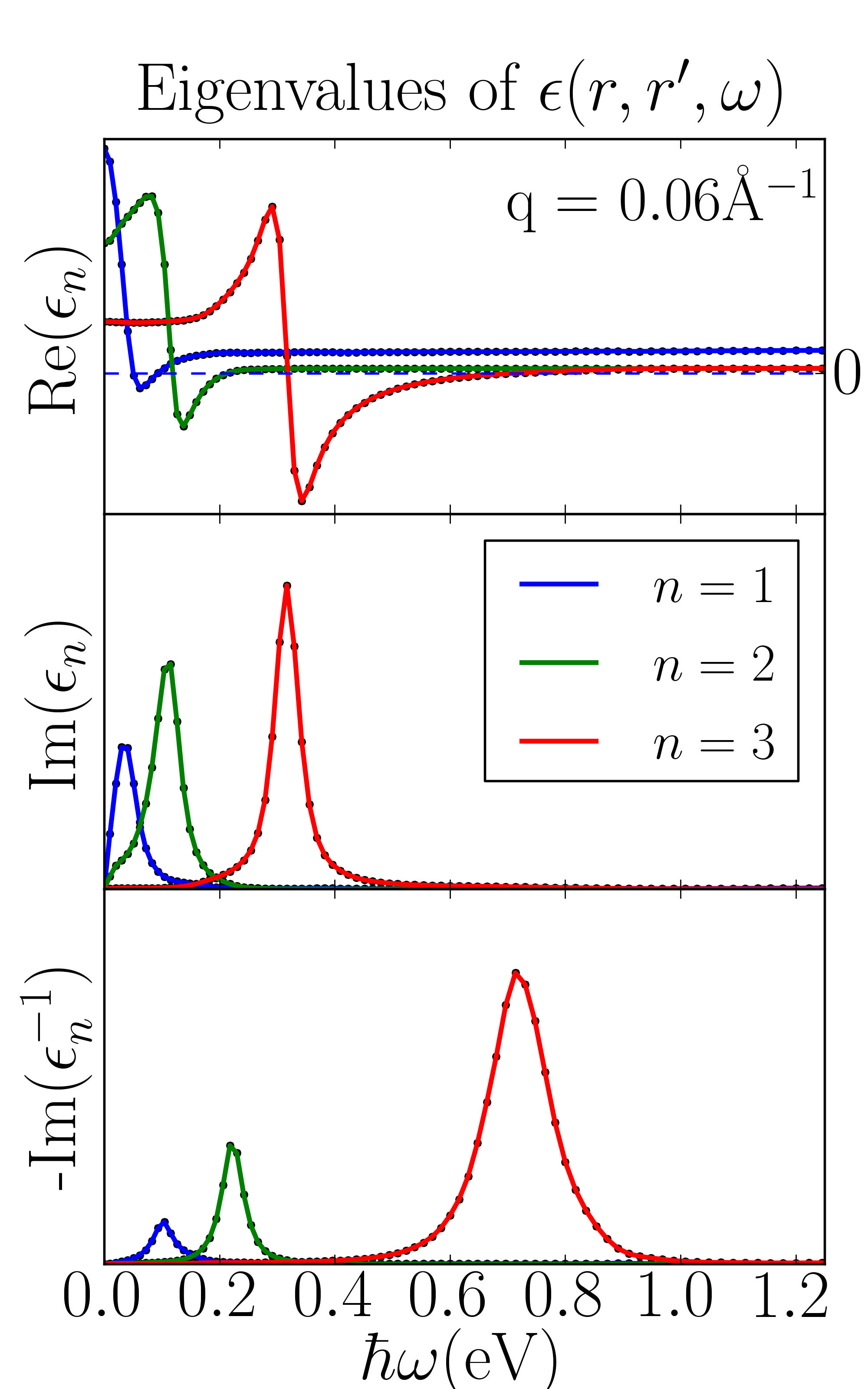}}
\caption{(a) 
Electronic bandstructure revealing three metallic edge-states (I-III). b) The three eigenvalues for the dielectric matrix fulfilling $\mathrm{Re} \, \epsilon_n(q,\omega)=0$ plotted for $q = 0.06 \mathrm{\AA}^{-1}$. The color (red,blue,green) illustrates the one to one connection between the electronic bands and the plasmon eigenmodes.}
\label{bands}
\end{figure}

The calculated band structure of the nanoribbon shown in Fig.~\ref{bands} reveals three metallic edge-states (labeled I-III) in agreement with previous calculations on a $\mathrm{MoS}_2$ nanoribbon with S-terminated edges \cite{bollinger2003atomic}. The remaining bands produce a band-gap of approximately 2 eV. When the width of the ribbon is doubled to 3.3 nm the edge states are unaffected while the bandgap is decreased to 1.8 eV which agrees with the value we find for the infinite single-layer $\mathrm{MoS}_2$. The Kohn-Sham wavefunctions I and III are almost completely localized on the S-dimers on the S- and Mo- edge of the ribbon respectively. The wavefunctions are linear combinations of $\mathrm{p}_x$-orbitals, forming the one-dimensional metallic states along the edge. (III) has a large Fermi-velocity and is close to linear up to 1 eV away from the Fermi level. The wave function of state II resides at the outermost Mo-atom at the Mo-edge, and extends a few atoms into the ribbon. The number of edge states and their properties depends on the particular edge-configuration. In this case the presence of sulfur dimers on the Mo-edge is crucial for obtaining the state (III). However, zigzag MoS$_2$ nanoribbons are in general found to be metallic opposed to armchair types \cite{pan2012edge}. We note that MoS$_2$ nanostructures could also have topologically protected metallic edge states due to strong spin-orbit coupling in MoS$_2$, which was not included in the present study. In the following we show that the metallic states give rise to plasmon states at the ribbon edges.

\begin{figure*}
\centering
  \subfloat[\label{dispa}]{\includegraphics[width=0.49\linewidth]{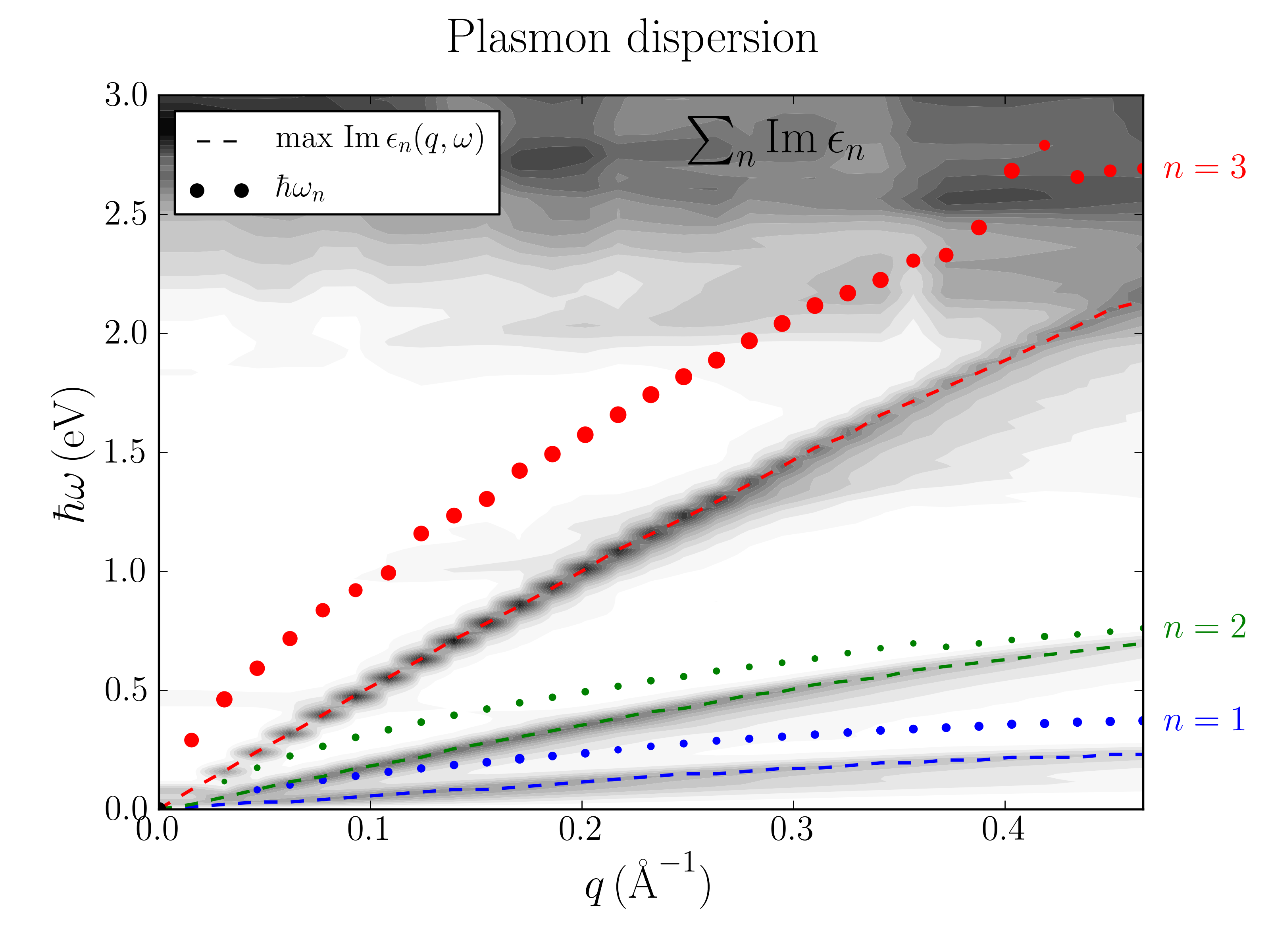}}
   \subfloat[\label{dispb}]{\includegraphics[width=0.49\linewidth]{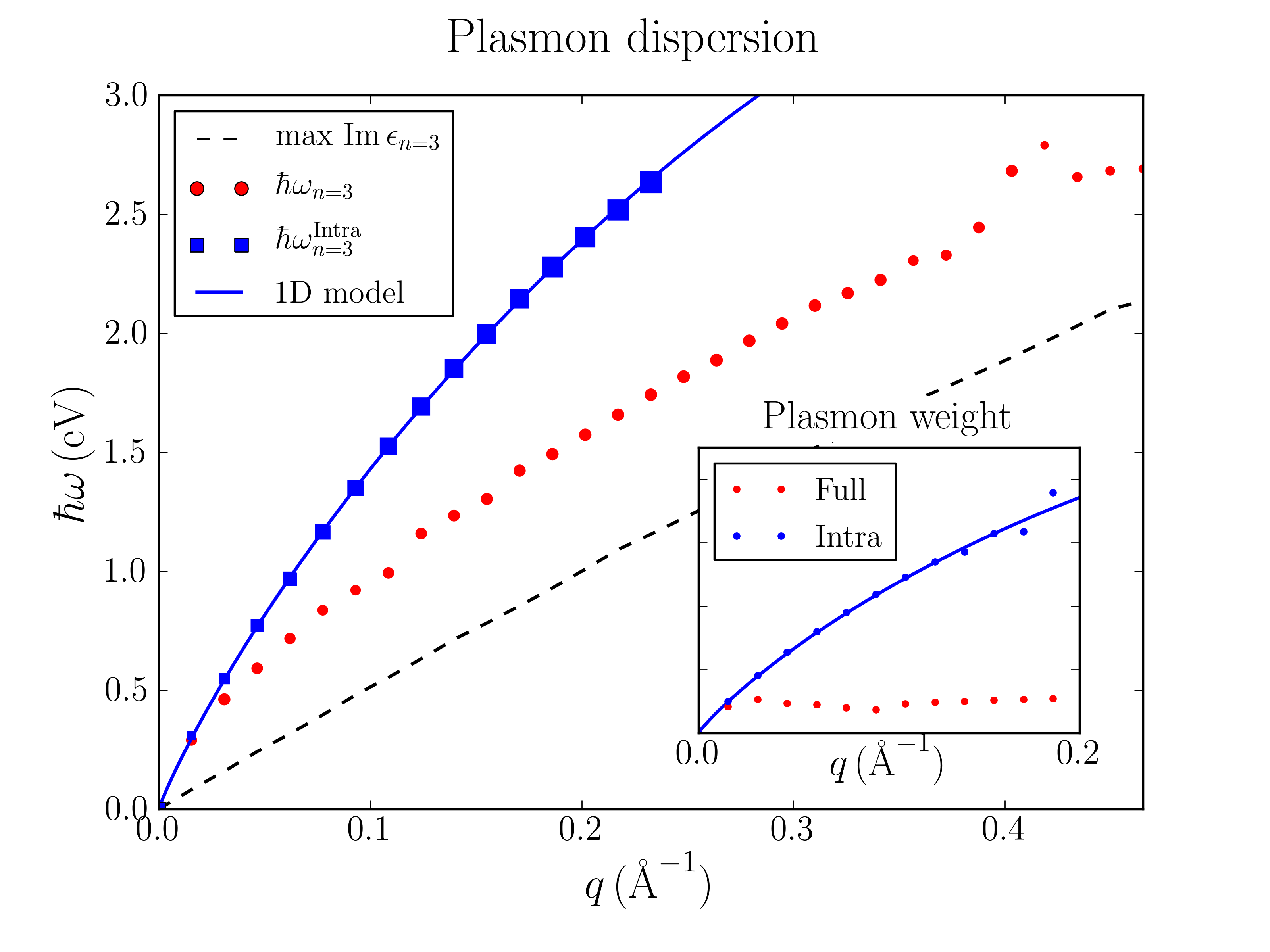}}
\caption{(a)Energy dispersion of the three edge plasmons, showing plasmon energy (dots) as well as intra-band transitions (dashed lines). The dot size indicates the plasmon weight, defined as the integral over the loss peak. The gray contours map the regions of single-particle transitions, that contribute to damping of the plasmons. (b) Comparison of the dispersion of the S-dimer plasmon to the undamped intra-band mode obtained by only including the metallic state (III) in the response calculation. The inset shows the evolution of plasmon weight with q for the undamped and the full calculation. A 1D RPA model valid at low $q$, identical to the Tamonaga-Luttinger result, has been fitted to the intraband results.}
\label{disp}
\end{figure*}

The plasmon modes of the system were investigated following a recently developed spectral analysis method for the dielectric function \cite{Andersen_2012}. The eigenvalue equation 
\begin{equation}
\sum_{\G'}\epsilon_{\G,\G'}(\mb{q},\omega) \phi^n_{\G'}(\mb{q},\omega) = \epsilon_n(q,\omega)\phi^n_{\G}(\q,\omega),
\end{equation}
is solved to obtain the frequency-dependent eigenvalues, $\epsilon_n$, and eigenvectors, $\phi^n$, of the microscopic dielectric function. The plasmon energies are identified as those frequencies where the real part of an eigenvalue vanishes: $\mathrm{Re}\, \epsilon_n(\mb{q},\omega_P) = 0$. When $\mr{Im} \, \epsilon_n(\mb{q},\omega)$ does not vary too much around the plasmon frequency, a vanishing real part coincide with a maximum in $-\mr{Im} \, \epsilon_n^{-1}(\mb{q},\omega)$.  The corresponding eigenvector $\phi^n(\mb{q},\omega_P)$ gives the spatial form of the induced potential associated with the plasmon oscillation. 

For the MoS$_2$ nanoribbon we find three distinct plasmon eigenmodes in the low-energy regime. The dielectric eigenvalue curves corresponding to these modes are shown in Fig.~\ref{eigs} for momentum transfer $q = 0.2 \mathrm{\AA}^{-1}$. Each mode arises from intraband transitions within one of the three metallic edge bands as indicated by the color code. The plasmons are clearly observed as peaks in $\mathrm{Im}\, \epsilon_n^{-1}(\omega)$ shown in the lower panel. In contrast, the imaginary parts, $\mathrm{Im}\, \epsilon_n$, have peaks at the energy of the intra-band transitions associated with the edge-states and occur at $~qv_F$.  The $n=3$ plasmon has a higher energy than the other two modes, partially due to the larger Fermi velocity of electrons in state (III). The induced potentials of the plasmon eigenmodes, given as the eigenvectors of the dielectic matrix, are shown in Fig.~\ref{modes}. The edge plasmon corresponding to $n=3$, is clearly localized on the S-dimers of the Mo-edge, and stems from the electronic state (III). The extend of the induced charge density of the edge plasmons are thus limited by the extend of the electronic edge states, so that the plasmons will be atomically localized independently of the size of the MoS$_2$ ribbon.   
 
In Fig.~\ref{dispa} the plasmon energies are plotted for increasing momentum transfers, and are clearly seen to be blue-shifted compared the the single-particle intra-band transitions (dashed lines) that disperse linear with $q$. In order to map out the effect of Landau damping in $(q,\omega)$-space, the sum of single-particle transitions, $\sum_{n} \mr{Im} \, \epsilon_n(q,\omega) = \mr{Im}\, \mr{Tr}\,\epsilon(\q,\omega)$, has been indicated in gray contours. The damping is seen become significant around 2 eV, where inter-band transitions between the valence and conduction band states of the MoS$_2$ sheet set in. 

At vanishing momentum transfers, the weight of single-particle transitions decreases significantly, since the number of single-particle transitions scales as $q$. For an ideal 1D electron gas the weight of the plasmon resonance should also go to zero in the limit $q \to 0$ \cite{sarma1996dynamical}. However, our calculations predict that the plasmon weight is almost constant in the considered $q$-range, and actually increases slightly for small momentum transfers. In order to explain this behavior, the result is compared to the pure intra-band plasmon, obtained by including only the single metallic band (III) (see Fig.~\ref{bands}) in the calculation of $\chi^0_{\G\G'}(\omega,\q)$. This allows us to compare to the case of an undamped plasmon. As seen in Fig.~\ref{dispb}, this approach yields a plasmon mode that is blue-shifted in energy compared to the full calculation and a weight that approaches zero for small $q$ as expected for an ideal 1D metal. Therefore, we conclude that the stagnant weight of the plasmon in the full calculation is due to damping from coupling to interband transitions. The effect of interband transitions can be described by an effective dielectric constant of the medium that screens the Coulomb interaction : $V(q) \to \epsilon^{-1}_{inter}(q,\omega) V(q)$. The screened interaction will be reduced in magnitude compared to $V(q)$ and will have a finite imaginary part at energies corresponding to the interband transitions. These properties lead to a reduction of energy and life time of the full plasmon compared to the bare intraband result. However, for very low momentum transfers, the plasmon is well separated from single-particle transitions, and a sharp resonance is obtained. Therefore the plasmon coincides with the undamped result, and is expected to have a long lifetime in this regime. For the other two  modes, we expect the damping by single-particle transitions to be lesssignificant due to separation in energy. 

Following reference~\cite{sarma1996dynamical}, a simple model for the plasmon dispersion in the long wavelength limit can be obtained by approximating $\chi_0$ as that of a 1D linear band with transitions at $q v_F$:
\begin{equation}
\chi_0(q,\omega) =  \frac{q}{\pi}\biggl[\frac{1}{\omega-q v_F + i\eta} - \frac{1}{\omega+q v_F+ i \eta}\biggr],
\end{equation}
The dielectric function is obtained from the RPA expression $\epsilon = 1-V(q)\chi_0$, where $V(q)$ is the 1D Coulomb potential. The inverse dielectric function can then be cast on the form
\begin{align}
&\epsilon^{-1}(q,\omega) = 1+ \frac{\alpha}{\omega-\omega(q)+ i \eta} -\frac{\alpha}{\omega+\omega(q)+ i \eta}. 
\end{align}
The pole of $\epsilon^{-1}$ for positive frequencies and in the limit of $\eta = 0$ yields the plasmon dispersion
\begin{equation}\label{eq:lut}
 \omega(q) = q v_F \sqrt{1+\frac{2 V(q)}{\pi v_F}},
\end{equation}
and the weight of the resonance equals $\alpha = q V(q)/\biggl(\pi \sqrt{1+\frac{2 V(q)}{\pi v_F}}\biggr)$. 
This result is identical to the eigen energies obtained from the interacting Tomonaga-Luttinger model with a 1D Coulomb interaction. In this sense the RPA is exact for a 1D system \cite{RPA_Lut1992}. The Coulomb interaction for a thin wire of confinement $a$ can be written for small $q$ as $V(q) = V_0 |\mathrm{ln}(qa)|$, where $V_0$ is a system dependent interaction strength accounting for e.g. the screening due to interband transitions. 
The expressions for the plasmon frequency and spectral weight derived from the linear band RPA model were fitted to the pure intraband results in Fig.~\ref{dispb}. As expected from the highly linear form of band (III), the quality of the fit is excellent. From the fit we obtained $a = 1.34 \mathrm{\AA}$ which is in the same order as the radius of the orbitals of the edge state. Comparing the dispersion and weight of the model to the results of the full ab initio calculation, we find agreement for small $(q,\omega)$, after which the plasmon energy and, in particular, the plasmon weight start to deviate in a nonlinear way. This shows that the screening varies as a function of $q$ and $\omega$ as is generally the case, such that the interaction strength cannot be written as a constant $V_0$ as assumed in the model. 
 \begin{figure}
\begin{center}
  \includegraphics[width=0.98\linewidth]{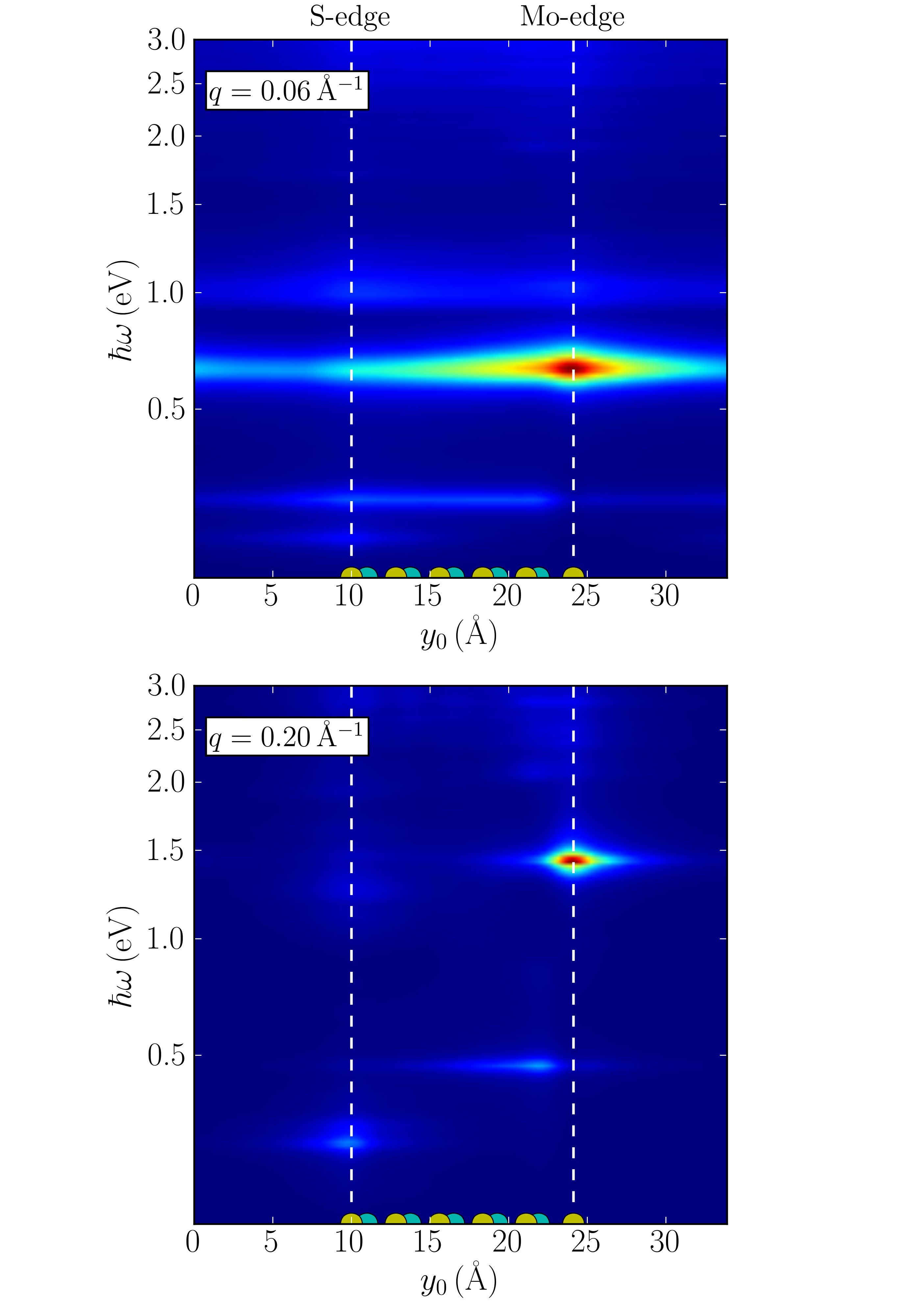}
\caption{Spatially resolved electron energy loss spectrum calculated for two different values of momentum transfer, $q_x$. We observe a strong signal associated with the S-dimer plasmon at the Mo-edge, while the two other plasmon modes give a weaker response. At low momentum transfers the signal is less localized due to the long range of the coulomb potential. }
\label{eels}
\end{center}
\end{figure}

Experimentally, plasmon excitations can be probed by electron energy loss spectroscopy (EELS). When performed with a highly confined electron beam, the spatial form of the loss spectrum can be obtained with few Angstrom resolution \cite{Abajo_2010}. 
 
The loss spectrum is defined as the power dissipated in the medium due to an external potential, $\phi_{ext}(\mb r)e^{i\omega t}$: 
\begin{align}\label{spateels}
P(\omega) =& \int \int d \mb r d \mb r' \phi_{ext}(\mb r) \chi_2(\mb r,\mb r',\omega) \phi_{ext}(\mb r').  
\end{align}
Here $\chi_2$ is the imaginary part of the interacting density response function. In the case of EELS, the external potential corresponds to the Coulomb potential of a fast electron moving at constant velocity $\ve$ emitted at point $\rr_0$: 
\begin{equation}
\phi_{ext}(\rr,t)= \frac{4 \pi e^2}{|\rr-\rr_0-\ve t |}. 
\end{equation}
The final expression for the loss spectrum is given in the supplemental material \cite{sup}. 
In Fig.~\ref{eels} the spatially resolved EELS spectrum is plotted for two different values of $q_x$ for a beam position that is varied acroos the ribbon. The spectrum is dominated by the S-dimer plasmon on the Mo-edge, which reflects the large coupling strength of this mode. The other two edge plasmons are also visible in the spectrum, particularly at larger momentum transfers where the Coulomb potential is less long ranged, which results in more localized features. Furthermore, the modes can be distinguished by their difference in energy and dispersion with $q$. We suggest this approach could be applied to verify the existence of the edge-plasmons in MoS$_2$ nanoribbons or clusters.

In conclusion, our first-principles calculations predict the existence
of a unique type of plasmon localized on the edge of a zigzag
$\mathrm{MoS}_2$ nanoribbon. The fundamental properties were investigated in detail and its clear signature in
the spatially resolved electron energy loss spectrum was demonstrated.
Finally, we proposed that the edge plasmons could be utilized to tune 
the photocatalytic activity of MoS$_2$ nanoparticles as was recently
demonstrated for metal nanoparticles \cite{marimuthu2013tuning, christopher2012singular}.
 
KST acknowledges support from the Danish Council for Independent Research's Sapere Aude Program through grant no. 11-1051390. The Center for Nanostructured Graphene (CNG) is sponsored by the Danish National Research Foundation, Project DNRF58.

\bibliography{bibtex}

\end{document}